\documentstyle[preprint,revtex]{aps}
\begin{document}
\draft
\begin{title}
\begin{center}
Transport Properties of the Infinite Dimensional Hubbard Model
\end{center}
\end{title}
\author{Th.\ Pruschke, D.L.\ Cox}
\begin{instit}
Department of Physics, The Ohio State University,\\
Columbus, Ohio 43210-1106
\end{instit}
\author{M.\ Jarrell}
\begin{instit}
Department of Physics, University of Cincinnati,\\
Cincinnati, Ohio 45221-0011
\end{instit}
\receipt{}
\begin{abstract}
Results for the optical conductivity and resistivity of the Hubbard
model in infinite spatial dimensions are presented. At half filling we
observe a gradual crossover from a normal Fermi-liquid with a Drude peak at
$\omega=0$ in the optical conductivity to an insulator as a function of
$U$ for temperatures above the antiferromagnetic phase transition.

When doped, the ``insulator'' becomes a Fermi-liquid with a corresponding
temperature dependence of the optical conductivity and resistivity. We find
a $T^2$-coefficient in the low temperature resistivity which suggests
that the carriers in the system acquire a considerable mass-enhancement
due to the strong local correlations. At high temperatures, a crossover into
a semi-metallic regime takes place.
\end{abstract}
\pacs{PACS numbers: 71.30.h, 72.15.-v, 72.20.-i}
\narrowtext
The one-band Hubbard Hamiltonian on a simple (hyper-)cubic lattice \cite{hbd}
\begin{equation}
H =-\sum_{\langle ij\rangle,\sigma}t_{ij} \left( c^\dagger_{i,\sigma}
c_{j,\sigma} + {\rm h.c.}\right)
-\mu\sum_{i\sigma}n_{i\sigma}
+U \sum_i n_{i,\uparrow}n_{i,\downarrow}\;\;,
\end{equation}
where $t_{ij}$ is restricted to nearest neighbours,
is from its structure one of the simplest models in condensed matter
physics. It is generally believed to contain the essential physics necessary
to qualitatively describe the properties of transition metal oxides and
possibly
the normal state of the high temperature superconductors.
Although a lot of work has been invested over the years \cite{voll-rev},
a reliable or even exact description of the physical properties of the Hubbard
model (1) for general values of its parameters could not be achieved except
for the limit of one spatial dimension \cite{1d}. Especially the interesting
issues of a possible metal-insulator transition or unconventional low
temperature
phases \cite{ffnj,and89} still remain unanswered.

A breakthrough towards a better understanding of highly correlated systems
in general, and the Hubbard model (1) in particular, was achieved by the
introduction of the limit of infinite spatial dimensions $d\to\infty$
\cite{mv89}. In this limit the dynamics
of the system become essentially local \cite{muha89-1}, which considerably
simplifies the task of calculating quantities of interest
\cite{bm,janis,ja92,gk,I,pcj}.
A particularly interesting result is that the one particle self energy
$\Sigma({\bf k},z)$
becomes purely local \cite{mv89}. Consequently one can map the thermodynamic
potential, and thus the solution of the model (1), onto an effective Anderson
impurity
problem \cite{bm,janis,ja92,gk}.
Recently, this method has been used by the present authors for a detailed study
of the phase diagram \cite{ja92,I} and thermodynamic and transport properties
\cite{pcj} of the Hubbard model using different approaches to solve the
underlying impurity problem \cite{pcj}.

In this letter we give a brief description of our results for the optical
conductivity and resistivity of the Hubbard Hamiltonian in the limit
$d\to\infty$. In particular, we want to address the abovementioned issues
of a possible metal-insulator transition and what kind of low temperature
behaviour the Hubbard model shows close to half filling and at realistic values
of $U$ in an artificially stabilized paramagnetic phase. It must be
stressed that the latter constraint is essential since the Hubbard model always
undergoes an antiferromagnetic transition close to half filling for
sufficiently low temperatures \cite{ja92,I}.
Before turning to a detailed discussion of our results let us give a brief
summary:
At half filling, one indeed encounters a crossover from a normal metal with a
Drude peak in the optical conductivity at small values of the Coulomb parameter
$U$ to an ``insulator'' with essentially zero dc-conductivity at large $U$. In
addition, a charge excitation peak at $\omega=U$ develops with increasing $U$
arising
from the field-induced transitions from the lower Hubbard band (LHB)
to the upper Hubbard band (UHB). Off half filling, we always find a
Fermi-liquid at low temperatures.
However, the corresponding Fermi-liquid parameter are strongly enhanced and
the resistivity shows nonmonotonic behaviour with a maximum at higher
temperatures.
The optical conductivity, on the other hand, has a
strongly temperature dependent Drude peak and
again the additional charge excitation peak at $\omega\approx U$.

In linear response theory the conductivity
can be expressed as
\FL\begin{equation}
\sigma(\omega)={\rm Re}\left\{\frac{1}{iz}\frac{1}{N}\sum_{{\bf k},{\bf
k}',\sigma}
\sum_lv_{k_l}v_{k'_l}\langle\langle n_{{\bf k}\sigma}|n_{{\bf k}'\sigma}
\rangle\rangle(z)\right\}_{z=\omega+i\delta}\;\;,
\end{equation}
where $v_{k_l}$ denotes the $l$-th component of the group velocity
${\bf v}_{\bf k}=\nabla\epsilon_{\bf k}$ of the carriers.
The expression in curly brackets in (4) has the perturbation expansion shown in
Fig.~1. For $d\to\infty$, momentum conservation at the irreducible vertex
part $\Gamma$ becomes unimportant, i.e.\ the two ${\bf k}$ sums
for the second and all higher order contributions in Fig.~1 can be performed
independently. Since ${\bf v}_{\bf k}$ and $\epsilon_{\bf k}$ have different
parity,
these contributions vanish identically \cite{khurana}, and we are left
with the simple bubble diagram. Inserting the usual tight-binding expression
$\epsilon_{\bf k}=-2t\sum\cos(k_la)$,
the bubble diagram can be evaluated using standard techniques and leads to the
final form
\widetext
\begin{equation}
\sigma(\omega)=\sigma_0\int d\omega'\int d\epsilon A_0(\epsilon)
A(\epsilon,\omega')A(\epsilon,\omega'+\omega)
\frac{f(\omega')-f(\omega'+\omega)}{\omega}
\end{equation}
\narrowtext
for the optical conductivity. Here, $A_0(\epsilon)=1/\sqrt{4\pi dt^2}
\exp[-\epsilon^2/(4dt^2)]$ denotes the one-particle density of states for the
noninteracting system, $A(\epsilon,\omega)$
the one with $U>0$, $f(\omega)$ is Fermi's function and $\sigma_0$ collects all
remaining constants and is given by
$$
\sigma_0=\frac{4d\pi e^2a^2t^2}{2\hbar}\frac{N}{Vol}\approx
10^{-2}\ldots10^{-3} [\mu\Omega cm]^{-1}
$$
for typical values $2\sqrt{d}t\approx 1{\rm eV}$ and $a=O(a_0)$. For
convenience
we shall choose $4dt^2=1$ in the following.

The only unknown quantity entering into (5) is the one-particle density of
states
$A(\epsilon,\omega)$ or equivalently the one-particle self energy
$\Sigma({\bf k},z)\stackrel{d\to\infty}{=}\Sigma(z)$.
As mentioned earlier, in $d=\infty$,
the task of determining $\Sigma(z)$ reduces to the solution of
an effective single impurity Anderson model \cite{ja92} and there exist several
reliable methods to handle this problem exactly by e.g.\ quantum Monte Carlo
methods \cite{ja92,I} or approximately within the so called NCA \cite{pg89}.
Because the quantum Monte Carlo does not provide the self-energy for real
frequencies,
the latter approach proves to be more adequate here.
We emphasize that the one-particle spectra obtained by the NCA
were carefully compared to the {\em exact} quantum Monte Carlo results
\cite{pcj} and found to be in excellent {\em quantitative} agreement over a
wide range of model parameters and temperatures. The results presented here
can thus be expected to be very close to the exact ones.

In Fig.~2 we show the variation of the one-particle density of states (Fig.~2a)
and the optical conductivity (Fig.~2b) for the half-filled case for various
values of $U$ calculated at a temperature $\beta=4$. The DOS
suggests a profound change in the qualitative behaviour of the system. Starting
from a single central peak at $\mu$ for $U\to0$, the characteristic
LHB and UHB
develop. The broad central peak, still present at intermediate $U$, gradually
vanishes
and is replaced by a pseudo-gap at $\mu$. This behaviour is reflected in the
optical conductivity in Fig.~2b. While for small $U$ only a Drude peak is
present,
the additional charge excitation peak at $\omega=U$ shows up with increasing
$U$. At
the same time the weight for $\omega\to0$ is drastically reduced. For $U=5$ we
find a situation one would expect for an insulator: Vanishing dc-conductivity
and a peak at $\omega=U$ reflecting these induced transitions from the LHB to
the UHB. The inset in Fig.~2b gives a logarithmic blow-up of the situation for
small $\omega$, clearly showing the reduction of the dc-conductivity by four
orders of magnitude as we go from $U=1$  to $U=5$.

Away from half filling, one important question is whether the Hubbard model
(1) is a conventional Fermi liquid or not? In Fig.~3 results for the DOS and
optical conductivity for some selected temperatures at fixed $U=4$ and
filling $n_e=0.97$ are shown. As for the DOS (Fig.~3a), we again observe the
temperature independent LHB and UHB separated by a pseudo-gap above $\mu$.
With decreasing temperature, however, a narrow quasiparticle band develops at
$\mu$. This band can be traced to a version of the Kondo-effect in this model
and one consequently has to expect a physical behaviour similar to the well
studied heavy fermion systems. In fact, the optical conductivity in Fig.~3b
shows a strongly temperature dependent Drude peak in addition to the charge
excitation peak at $\omega\approx U$. One also clearly sees that the spectral
weight
built up at $\omega\to0$ is taken from this peak at $\omega\approx U$. The
inset
gives again an enlarged view of the situation for $\omega\to0$.

It is of course interesting to study the temperature dependence of $\sigma(0)$
or $\rho(T)=1/\sigma(0)$ more closely. If the system were a Fermi-liquid at low
temperatures, we should observe $\rho(T)\sim(T/T_0)^2$, where $T_0$ is the
effective bandwidth for the quasiparticles. Figure 4 shows the resistivity
for $U=4$ and $n_e=0.97$ along with the low temperature data versus $T^2$
in the inset. The latter nicely fall onto the curve $\rho(T)=a\cdot(T/T_0)^2$,
where $a=O(1)$ and $T_0$ is roughly the width of the quasiparticle peak at
$\mu$
in the DOS in Fig.~3a. From this result one clearly has to conjecture that at
least for high spatial dimensions the low temperature (paramagnetic)
phase of the Hubbard model for fillings $n_e<1$
is a (possibly heavy) Fermi-liquid. At higher temperatures the
resistivity changes from a metallic to a semi-metallic behaviour. This is also
consistent with the vanishing of the quasiparticle peak at $\mu$ for high
temperatures.

To summarize, we presented characteristic results for the optical conductivity
and resistivity of the Hubbard model in $d\to\infty$. The data shown were
calculated within an approximate method to solve the impurity Anderson model,
which essentially determines the physics of the Hubbard model in this limit.
Let us stress that this approximation was shown to very accurate in its
domain of applicability \cite{pcj}. The results at half filling suggest a
crossover from a normal metal to an insulator as a function of $U$ with a
$U_c\approx3\ldots4$. Further results not presented here \cite{pcj} show
that this scenario does not change qualitatively for lower temperatures.
Off half filling we encounter a Fermi-liquid at low temperatures together
with a crossover to a semi-metal at high temperatures.
In both cases the optical conductivity reflects the relevant low temperature
energy scales of the model, the bare bandwidth $t$ at small $U$ and the
Coulomb energy plus possibly a dynamically generated scale $T_0$ for the
quasiparticles in the Fermi-liquid for large $U$.

The general behaviour of the two quantities studied here fits at least
qualitatively into the picture of the normal state
properties of high-$T_c$ superconductors. There are, though, obvious
differences, the most important is probably the missing linear low temperature
resistivity generally found in the cupratres.
One must, however, bear in mind that our results are strictly valid only for
$d\to\infty$
and that especially in 2D dimensional effects will of course be important. For
example, one aspect clearly absent in $d\to\infty$ are spin-fluctuations.
As was shown recently \cite{ueda-moriya} the
coupling to these degrees of freedom may easily change the power-law in
$\rho(T)$. Thus, with a proper account for corrections due to finite
dimensions, our
results may well be a first step towards the solution of outstanding questions
in the field of highly correlated systems.

\acknowledgements
This research was supported by the National Science Foundation grant number
DMR-88357341, the National Science Foundation grant number DMR-9107563, the
Ohio State University center of materials research and by the Ohio
Supercomputing
Center.

\end{document}